\documentclass[a4paper,11pt] {article}
\newcommand{\be}{\begin{equation}}
\newcommand{\ee}{\end{equation}}
\newcommand{\ba}{\begin{eqnarray}}
\newcommand{\ea}{\end{eqnarray}}
\newcommand{\bea}{\begin{eqnarray}}
\newcommand{\eea}{\end{eqnarray}}

\begin{document}

\title{Solvable few-body 
quantum problems  }
\author{ A. Bachkhaznadji \\
Laboratoire de Physique Th\'eorique \\
D\'epartement de Physique \\
Universit\'e  Mentouri \\
Constantine, Algeria\\
M. Lassaut   \\
Institut de Physique Nucl\'eaire\\
 IN2P3-CNRS and Universit\'e  Paris-Sud   \\
F-91406 Orsay Cedex,  France \\ [3mm]
\date{\today }}
\maketitle

Abstract :

This work is devoted to the study of  some exactly solvable quantum  problems of four, five and six bodies moving on the line. We solve completely the corresponding stationary Schr\"odinger equation for these systems confined  in an harmonic trap, and  interacting pairwise, in clusters of two and three particles, by two-body inverse square Calogero
 potential. Both translationaly and non-translationaly invariant multi-body potentials are added.
 In each case,   the full solutions are provided, namely the normalized regular eigensolutions and the
 eigenenergies spectrum.
  The irregular solutions are also studied.   We discuss the domains of  coupling   constants for which these irregular solutions are square integrable. The case of a "Coulomb-type" confinement is investigated only for the bound states of the four-body systems.

  
\newpage

\section{Introduction}

A limited number of exactly solvable many-body systems  exists,  in the one dimensional space  \cite{mattis,sutherland}.
The Calogero model constitutes  a famous example, which was
exhaustively studied \cite{Calo69p,Calo71}.
A survey of quantum integrable systems was done by 
Olshanetsky and Perelomov  \cite{Perelomov1983}.
They classified the systems with respect to Lie algebras.
Point interactions have also been considered, still in $D=1$
dimensional space \cite{albe1,albe2}.

The search  for exactly solvable few-body systems concerns 
 mostly the three-body case and  still retains attention.
Early works of three-body linear problems of
 Calogero-Marchioro-Wolfes  \cite{Calo69,CM74,Wolfes74}
have been followed by new extensions and cases.
In a non exhaustive way, we can quote the three-body 
problems of Khare {\it et al.} \cite{KB}, Quesne \cite{ques2}
and Meljanac  {\it et al.} \cite{Meljanac2007}.
In a recent paper, we have exactly solved a generalization of the three-body
Calogero-Marchioro-Wolfes problem
with an additional non-translationally three-body potential \cite{BLL2009}. 

Works on the four-body problem and beyond   are much more scarce \cite{Wolf74,chin}.
This is due to the difficulty to extend right away the Calogero model to a larger number of particles.
We can quote  the work of Haschke and  R\"uhl \cite{Has98} concerning  the
construction of exactly solvable quantum models of Calogero and Sutherland type
with translationally invariant two-and four particles interactions.

Recently   an extension of  the four-body problem of Wolfes \cite{Wolf74}   with  non-translationally 
invariant interactions   has been investigated and solved  exactly \cite{BL13}.  
The purpose of the present work is to point out particular cases admitting an exact solution. In the four-body problem,
we consider  the particles to be experiencing either  an harmonic field or to be  submitted  to an attractive $1/r$-type potential. The interaction among the particles is inspired by the one used in our previous work on the three-body potential \cite{BLL2009}. It contains two- and  many-body forces.

The paper is organized   as follows.
In section {\bf 2} we expose and solve the four-body problem
 for the case of both harmonic and "Coulomb-type" confinement of the particles.
 Section {\bf 3} and {\bf 4} are devoted to the solutions of similar five- and six-body problems respectively.
 Conclusions are drawn in section {\bf 5}.

\section{A four-body  problem}

\subsection{The case  with  harmonic confinement.}

We consider the Hamiltonian
\begin{equation}
H=\sum_{i=1}^{4}\left( -\frac{\partial ^{2}}{\partial x_{i}^{2}}%
+\omega ^{2}x_{i}^{2}\right) +\lambda \sum_{i<j}^3 \frac{1}{%
(x_{i}-x_{j})^{2}}+\frac{g}{(x_1+x_2+x_3- 3  x_4)^2} +\frac{\mu }{\sum_{i=1}^{4}x_{i}^{2}}
%\text{ ,}
\label{3c-1}
\end{equation}%
Here, we use the units $\hbar=2 m =1$.  The first contribution gives the energy of four 
independent particles in an harmonic field. 
To this part, residual interactions are added in the following way.
  The  first three particles, with coordinates $x_1,x_2$ and $x_3$,   interact pairwise  by a two-body inverse square potentials, 
of Calogero type \cite{Calo69}.
 Their  centre of mass interacts  with the fourth particle, whose  the coordinate is  $x_4$,  via the  translationally invariant  term  $g/(x_1+x_2+x_3- 3  x_4)^2$. 
 An additional non translationally invariant  four-body  potential, represented by the term 
$\mu/(x_1^2+x_2^2+x_3^2+x_4^2)$, is added to the whole Hamiltonian.

In order to solve this four-body problem,
let us introduce the following coordinates transformation 
\begin{equation}
 t=\frac{1}{2}(x_{1}+x_{2}+x_{3}+x_4),\quad u=\frac{1}{\sqrt{2}}(x_{1}-x_{2}),\quad v=\frac{1}{\sqrt{6}}(x_{1}+x_{2}-2x_{3}),\quad w=\frac{1}{2 \sqrt{3}} (x_1+x_2+x_3-3 x_4).
\label{3c-16bis}
\end{equation}
The transformed Hamiltonian reads:
\begin{eqnarray}
H &=&-\frac{\partial ^{2}}{\partial t^{2}}-\frac{\partial ^{2}}{\partial w^{2}}-\frac{%
\partial ^{2}}{\partial u^{2}}-\frac{\partial ^{2}}{\partial v^{2}}+\omega
^{2}(t^{2}+u^{2}+v^{2}+w^2)    +\frac{\mu }{t^{2}+u^{2}+v^{2}+w^2}  \nonumber \\ 
& & +\frac{9\lambda (u^{2}+v^{2})^2}{%
2\left( u^{3}-3uv^{2}\right) ^{2}}  +\frac{g }{12 w^2}.  \label{3c-48}
\end{eqnarray}
This Hamiltonian is not  separable in  $\{t,u,v,w\}$ variables.
To overcome this situation we introduce the following  hyperspherical coordinates : 
\begin{eqnarray}
t &=&r\cos \alpha, \quad w=r\sin \alpha \cos \theta, \quad u=r\sin \alpha \sin \theta \sin \varphi,
\quad v=r\sin \alpha \sin \theta \cos \varphi ,\quad  \nonumber \\
0 &\leq &r<\infty ,\quad \quad  0\leq \alpha \leq \pi ,\quad\quad\quad\quad   0\leq \theta \leq \pi,\quad \quad\quad\quad\quad 0\leq
\varphi \leq 2\pi .  \label{3c-18bis}
\end{eqnarray}

The Schr\"{o}dinger equation is then written:
\begin{eqnarray}
&&\left\{ -\frac{\partial ^{2}}{\partial r^{2}}-\frac{3}{r}\frac{\partial }{%
\partial r}+\omega ^{2}r^{2}+\frac{\mu }{r^{2}}  +  \frac{1}{r^{2}}\left[
 -\frac{\partial ^{2}}{\partial \alpha ^{2}}   -2 \cot \alpha \frac{\partial }{\partial \alpha } + 
\frac{1}{\sin ^{2}\alpha}  \left(  -\frac{\partial ^{2}}{\partial \theta ^{2}}-\cot \theta \frac{\partial }{\partial \theta }
 \right.\right. \right. \nonumber\\
 & & \left.\left.\left. + \frac{g}{12 \cos ^{2}\theta }+ \frac{1}{\sin ^{2}\theta }\left(- 
\frac{\partial ^{2}}{\partial \varphi ^{2}}+\frac{9\lambda }{2\sin
^{2}(3\varphi )}\right) \right) \right] \right\} \Psi(r,\alpha,\theta,\varphi)  =E\Psi(r,\alpha,\theta,\varphi)   \ .  \label{3c-49}
\end{eqnarray}%

This Hamiltonian may be mapped to
 the problem of one particle in the four dimensional space with a non central potential of the form
\begin{equation}
V(r,\alpha,\theta ,\varphi )=f_{1}(r)+\frac{1}{r^{2}\sin ^{2}\alpha } \left[
f_{2}(\theta )+ \frac{f_{3}(\varphi )}{\sin ^{2}\theta } \right].
\end{equation}
It is then clear that the problem becomes separable in the  four variables $\{r,\alpha,\theta,\varphi\}$.
To find the solution we factorize the wave function as follows :
\begin{equation}
\Psi_{k,\ell,m,n} (r,\alpha,\theta ,\varphi )=\frac{F_{k,\ell,m,n}(r)}{r \sqrt{r}} \frac{G_{\ell,m,n}(\alpha)}{\sin \alpha}
 \frac{\Theta_{m,n} (\theta )}{\sqrt{\sin
\theta }}\Phi_n (\varphi ).  \label{3c-6} 
\end{equation}
Accordingly,  equation (\ref{3c-49}) separates in  four   decoupled  differential equations:
\begin{equation}
\left( -\frac{d^{2}}{d\varphi ^{2}}+\frac{9\lambda }{2\sin^{2}(3\varphi )}%
\right) \Phi_n (\varphi )=B_n\Phi_n (\varphi ),\quad  \label{3c-7}
\end{equation}
\begin{equation}
\left( -\frac{d^{2}}{d\theta ^{2}}+\frac{(B_n-\frac{1}{4})}{\sin ^{2}\theta }%
+\frac{g}{12 \cos^2\theta} \right) \Theta_{m,n} (\theta )=C_{m,n} \Theta_{m,n} (\theta ),\qquad  \label{3c-8}
\end{equation}%
\begin{equation}
\left( -\frac{d^{2}}{d\alpha ^{2}}+\frac{C_{m,n}-\frac{1}{4}}{\sin ^{2}\alpha }%
\right) G_{\ell,m,n} (\alpha )=D_{\ell,m,n} G_{\ell,m,n} (\alpha ),\qquad  \label{3c-9}
\end{equation}%
and 
\begin{equation}
\left( -\frac{d^{2}}{dr^{2}}+\omega ^{2}r^{2}+\frac{\mu +D_{\ell,m,n}-\frac{1}{4}}{r^{2}%
}\right) F_{k,\ell,m,n}(r)=E_{k,\ell,m,n}\  F_{k,\ell,m,n}(r) \ .  \label{3c-10}
\end{equation}

 The potential of Eq.(\ref{3c-7})  has a periodicity of  $\frac{\pi }{3}$.
In the interval $0\leq \varphi \leq 2\pi $, it   possesses singularities at
 $\varphi =k \frac{\pi }{3},k=0,1,...,5 $. 
This equation   has been solved by Calogero \cite{Calo69}.
In  the vicinity of  $\varphi=0$ (resp. $\frac{\pi }{3},$) the singularity
can be treated if and only if $\lambda >-1/2$, similar to the case
 of a centrifugal barrier.  Otherwise the operator has several self-adjoint extensions, 
each of which may lead to a different spectrum \cite{Zno,Barry}.

The solutions of equation (\ref{3c-7})  on the interval $[0,\pi/3]$,  with Dirichlet conditions at the boundaries, in terms 
 of orthogonal polynomials, are given by  \cite{Calo69,BLL2009}
\begin{equation}
\Phi _{n}^{(\pm)}(\varphi ) = (\sin 3\varphi )^{\frac{1}{2} \pm a}C_{n}^{(\frac{1}{2}
\pm a)}(\cos 3\varphi ),     \quad 0\leq \varphi \leq \frac{\pi }{3},    
          \qquad n=0,1,2,...
\label{3c-22} 
\end{equation}
The $C_n^{(q)}$ denote the Gegenbauer polynomials \cite{erd}.
The corresponding eigenvalues take the form
\begin{equation}
B_{n}^{(\pm)} =9\left( n  + \frac{1}{2} \pm a \right) ^{2},\qquad
n=0,1,2,....  \label{3c-22-bis}
\end{equation}
%In equations (\ref{3c-22},\ref{3c-22-bis}) 
 with
\begin{equation}
a =\frac{1}{2} \sqrt{1+2\lambda } \ .
\label{symba}
\end{equation}

The extension  to the whole interval $[0,2 \pi] $ is achieved  following the prescription given in \cite{Calo69},
by  using symmetry arguments according to the statistics obeyed by  the particles.
 Generally, only the regular  solution, $\Phi _{n}^{(+)}, $ corresponding  to $1/2+a$, is retained.
However,  the irregular solution,  $\Phi _{n}^{(-)} $, which is distinct of $\Phi _{n}^{(+)} $ for
$\lambda >-1/2$, is physically  acceptable when the  Dirichlet condition is satisfied for $-1/2 < \lambda \leq 0$ 
(attractive potentials). 
 If we release the Dirichlet condition, and ask only for the square integrability of the solution,  as in \cite{Murthy92}, 
then $\Phi _{n}^{(-)}$ can be retained for $-1/2 <\lambda <3/2$ \cite{BLL2009}.
For $\lambda=0$,
 the pairwise interaction between the particles $1,2$ and $3$ disappears.

The second angular equation for the polar angle $\theta$ can be written as
\begin{equation}
\left( -\frac{d^{2}}{d\theta ^{2}}+\frac{b_{n}^{2}-\frac{1}{4}}{\sin
^{2}\theta }+\frac{\beta}{\cos ^2\theta} -C_{m,n}\right) \Theta_{m,n} (\theta )=0,  \label{3c-24}
\end{equation}%
where the auxiliary constants  $b_{n }, \beta$  are defined respectively by
\begin{equation}
b_{n}^{2}=B_{n},\qquad b_{n}=\sqrt{B_{n}}=\left( 3n+3a+\frac{3}{2} \right).  \label{3c-21}
\end{equation}
and
\begin{equation}
\beta=\frac{g}{12} \ .  \label{3c-21b}
\end{equation}
Since  $b_n \ne  0$, the Hamiltonian of Eq.(\ref{3c-24}) is a self-adjoint operator
 provided that $\beta >-1/4$.
 
 Considering the interval $0 < \theta < \pi $,
 we first note that Eq.(\ref{3c-24}) has three singularities 
 occurring at $\theta = k\frac{\pi }{2}$, with $k=0,1,2.$ 
This separates the interval $]0,\pi[$ in two equal length intervals namely 
\begin{eqnarray}
0 &<& \theta <\frac{\pi}{2} \quad  {\rm with}  \quad r \sin \alpha \cos \theta
=w >0, \label{wp} \qquad \\
\frac{\pi}{2} &<&\theta <\pi \quad {\rm with} \quad r \sin \alpha \cos \theta
=w <0 \ .
\label{wm}
\end{eqnarray}
It corresponds  to a positive (or negative) value  of  $w$, respectively.
 We remind the reader that $\sin(\alpha) > 0$ as $\alpha \in ]0,\pi[$.
In the interval $\theta \in ]0,\pi/2[$ (resp.  $\theta \in ]\pi/2,\pi[$)
the centre of mass  of particles $1,2$ and $3$  is situated at the right 
 (resp. the left) of the fourth particle. 
The presence of the centrifugal term in $\beta/\cos(\theta)^2$ 
 forbids the c.m coordinate of particles $(1,2,3)$ to coincide with $x_4$.

The solutions of Eq.(\ref{3c-24})  in $]0,\pi/2[$ with Dirichlet conditions
 are well known \cite{BLL2009}. 
The regular solutions are
 \begin{eqnarray}
\Theta^{(+)} _{m,n}(\theta ) &=&(\sin \theta )^{b_{n}+\frac{1}{2}}(\cos \theta )^{c+%
\frac{1}{2}}P_{m}^{(b_{n},c)}(\cos 2\theta ),\qquad  \label{3g-17} \\
\quad 0 &\leq &\theta \leq \frac{\pi }{2},\quad m=0,1,2,...,\quad c=\frac{1}{%
2}\sqrt{1+4\beta }=\frac{1}{2} \sqrt{1+\frac{g}{3} }     \  .      
\label{thetap}
\end{eqnarray}
 Here,  $P_{m}^{(b_{n},c)}$ are the Jacobi polynomials  \cite{erd}.
The index $+$ means that the 
$w$-coordinate is positive. The solutions in the  interval $\frac{\pi }{2} < \theta <  \,\pi $ are 
\begin{eqnarray}
\Theta^{(-)} _{m,n}(\theta) &=&(\sin \theta
)^{b_{n}+\frac{1}{2}}(-\cos \theta )^{c+\frac{1}{2}}P_{m}^{(b_{n},c)}(\cos
2\theta ), \label{3gp-17} \qquad \\
\frac{\pi}{2} & < &\theta < \pi ,\quad m=0,1,2,...\,.  \nonumber
\end{eqnarray}
The index $-$ means the $w$-coordinate to be  negative.
Note that in the interval $]\pi/2,\pi[$ we have $-\cos \theta > 0 $ and $\sin \theta > 0$ so that 
the real power of these positive values are defined.

In both cases, the eigenvalues  $C_{m,n}$ of Eq.(\ref{3c-24}) are given by 
\begin{equation}
C_{m,n}=(2 m +b_{n}+c+1)^{2},\quad m=0,1,2,...\,.\quad
\label{Cmn}
\end{equation}

The third angular equation (\ref{3c-9}) has been treated in \cite{BLL2009}.
The regular eigensolutions and corresponding eigenvalues in the interval $]0,\pi[$
read, respectively, 
\begin{eqnarray}
G _{\ell,m,n}(\alpha ) &=&(\sin \alpha )^{c_{m,n}+\frac{1}{2}}C_{\ell}^{(c_{m,n}+%
\frac{1}{2})}(\cos \alpha ),\qquad \ell=0,1,2,...,\quad  \label{3c-30} \\
D_{\ell,m,n} &=&\left( \ell+c_{m,n}+\frac{1}{2} \right)^{2},\qquad \qquad \qquad \,\, \ell=0,1,2,...\ .
\label{3c-31}
\end{eqnarray}
 Here, $c_{m,n}=\sqrt{C_{m,n}}$ (see Eq.(\ref{Cmn})).
Note that the choice  $c_{m,n} >0$ implies  for every value of  $\{m,n\}$ the function
$G_{\ell,m,n}(\alpha)$  to vanish at the boundaries  of the interval  $\left ]0,\pi \right[.$

Finally, the reduced radial equation reads 
\begin{equation}
\left( -\frac{d^{2}}{dr^{2}}+\omega ^{2}r^{2}+\frac{\mu +D_{\ell,m,n}-\frac{1}{4}}{%
r^{2}}-E_{k,\ell,m,n} \right) F_{k,\ell,m,n}(r)=0.  \label{3c-32}
\end{equation}%
This is nothing but the usual 3-dimensional harmonic oscillator equation,
 $(\mu + D_{\ell,m,n}-1/4)/r^2$ replacing the centrifugal barrier. The 
 square integrable solutions are well known, putting a limit on the coefficient 
 of the $1/r^2$ term.
%It is solved in the interval  $0\leq r<\infty ,$ with the condition of square integrability for the solutions.
%It implies $F_{k,\ell,m,n}(r) \to 0$ as $r \to \infty$.
%We have to impose $\mu + D_{\ell,m,n} >3/4$ in order to treat the centrifugal barrier in the vicinity of $r=0$.
Note that taking $\mu + D_{\ell,m,n} =0$ leads to several self-adjoint extensions differing by a phase.
This fact has been discussed in \cite{BLL2009}. More details can be found in \cite{basu,giri}.
It has to be noted that for attractive centrifugal barriers,  $\mu + D_{\ell,m,n} <0$, the problem of collapse appears, unless
 regularization procedures are taken into account \cite{case,gupta,camblong,YLL2013}.
 Using the definition of $D_{\ell,m,n}$,  Eq.(\ref{3c-31}), we have
\begin{equation}
\mu +D_{\ell,m,n}=\mu +\left(\ell+2 m + 3 n + c + 3  a + 3 \right)^{2}> 0, \quad
\forall n\geq 0,\quad \forall m\geq 0, \quad \forall \ell \geq 0.  \label{3c-32bisp}
\end{equation}
The quantity $\mu +D_{\ell,m,n}$ is minimal for  $n=0,m=0,\ell=0$ and $a=c=0$ ( we recall that $a \geq 0$, see (\ref{symba})
 and that $c \geq 0$, see (\ref{thetap})). 
It put constraint on  $\mu$, which has to satisfy  $\mu >-9$.  
 We introduce the auxiliary parameter  $\kappa _{\ell,m,n}$ defined by
\begin{equation}
\kappa _{\ell,m,n}^{2}=\mu +D_{\ell,m,n}, \qquad \kappa _{\ell,m,n}=
\sqrt{\mu +D_{\ell,m,n} \ }.
\label{3c-33}
\end{equation}
 The solution of the radial equation (\ref{3c-32})  vanishing at $r=0$ and for $r \to \infty$ is

\begin{equation}
F_{k,\ell,m,n}(r)=r^{\kappa _{\ell,m,n}+\frac{1}{2}}\exp \left(-\frac{\omega r^{2}}{2}
\right)L_{k}^{(\kappa _{\ell,m,n})}(\omega r^{2}),\qquad k=0,1,2... \ .  \label{3c-37}
\end{equation}
The $L_k^{(q)}$ are the generalized Laguerre polynomials \cite{erd}.
The associated eigenenergies are given by
\begin{equation}
E_{k,\ell,m,n}=2\omega (2k+\kappa _{\ell,m,n}+1),\qquad k=0,1,2....  \label{3c-38}
\end{equation}

Collecting all pieces, we conclude  the physically acceptable solutions of 
the Schr\"odinger equation  (\ref{3c-49}) to be given by  
\begin{eqnarray}
\Psi _{k,\ell,m,n}(r,\alpha,\theta ,\varphi ) &=&r^{\sqrt{\mu +(\ell+ 2m +3n+c +3a+3)^{2} \ } \ -
1}e^{-\frac{\omega r^{2}}{2}}L_{k}^{\left( \sqrt{\mu +(\ell+2m +3n+c+3a+2)^{2} \  }
\right) }(\omega r^{2})  \nonumber \\
&&\times (\sin \alpha )^{2m +3n+c+3a+2}C_{\ell}^{^{\left( 2m+3n+c+3a+3 \right)
}}(\cos \alpha )  \nonumber \\
&&\times (\sin \theta )^{3n+3a+\frac{3}{2}} (\epsilon_1 \cos \theta)^{c+\frac{1}{2}} P_{m}^{\left( 3n+3a+\frac{3}{2},c \right)
}(\cos 2 \theta )  \nonumber \\
&&\times (\sin 3\varphi )^{a+\frac{1}{2}}C_{n}^{\left( a+\frac{1}{2}\right)
}(\cos 3\varphi ),\,  \label{3c-39pp} \\
k &=&0,1,2,...,\qquad \ell=0,1,2,...,\qquad m=0,1,2,...,,\qquad n=0,1,2,...,   \nonumber \\
 0 \leq \varphi \leq \frac{\pi }{3}\ , \quad & & \frac{1-\epsilon_1}{2} \frac{\pi}{2}  \leq \theta \leq 
\frac{3-\epsilon_1}{2} \frac{\pi }{2} , \epsilon_1=\pm 1,\quad a=\frac{1}{2}\sqrt{
1+2\lambda } \ ,\quad c=\frac{1}{2}\sqrt{1 + \frac{g}{3} \ } \quad \quad \nonumber    
\end{eqnarray}
Using the prescription of \cite{Calo69}
and the parity properties of the Gegenbauer polynomials \cite{erd},
we can write the solution as a  compact form,   valid in the whole interval $[0,2 \pi]$ for $\varphi$ 
\begin{eqnarray}
\Psi _{k,\ell,m,n}(r,\alpha,\theta ,\varphi ) &=&r^{\sqrt{\mu +(\ell+ 2m +3n+c +3a+3)^{2} \ } \ -
1}e^{-\frac{\omega r^{2}}{2}}L_{k}^{\left( \sqrt{\mu +(\ell+2m +3n+c+3a+2)^{2} \ }
\right) }(\omega r^{2})  \nonumber \\
&&\times (\sin \alpha )^{2m +3n+c+3a+2}C_{\ell}^{^{\left( 2m+3n+c+3a+3 \right)
}}(\cos \alpha )  \nonumber \\
&&\times (\sin \theta )^{3n+3a+\frac{3}{2}} ( \epsilon_1 \cos \theta )^{c+\frac{1}{2}} P_{m}^{\left( 3n+3a+\frac{3}{2},c \right)
}(\cos 2 \theta )  \nonumber \\
&&\times sgn(\sin 3\varphi)^{[(1-\epsilon_2)/2]}   \vert \sin 3\varphi \vert^{a+\frac{1}{2}}C_{n}^{\left( a+\frac{1}{2}\right)
}(\cos 3\varphi ) \ .   
\label{comp}
\end{eqnarray}
Here, $\epsilon_2=1$ for bosons and $-1$ for fermions in equation Eq.(\ref{comp}).
 We recall that   $sgn(x)=x/\vert x \vert$ denotes the sign of  $x \ne 0$.  
For the Bose statistics,  the extension  (\ref{comp}) is possible  provided that    no $\delta$ distribution 
occurs  when the second derivative of the wave function with respect to $\varphi$
is applied  at the boundaries  between two adjacent sectors.
For example,  for  $\varphi=\pi/3$ and $n=0$ , a $\delta$ distribution occurs for 
$a=1/2$ implying  $\lambda=0$. 
 It is due to the presence of $\vert \sin 3 \varphi \vert$ in (\ref{comp}). 
As a consequence,  the symmetrical solutions 
 for the pure harmonic oscillator ($\lambda=\mu=g=0$) are not recovered.
Also a $\delta$ distribution occurs for $c=1/2$ i.e. $g=0$ and we do not recover the
 pure solutions $g=0$.

The normalization constants $N_{k,\ell,m,n}$ can be  calculated from
\begin{eqnarray}
&& \int_{0}^{+\infty }r^{3}dr\int_{0}^{\pi }\sin^2 \alpha \  d\alpha 
\int_{(1-\epsilon_1) \pi/4}^{(3-\epsilon_1) \pi/4} \sin \theta \ d\theta \int_0^{\frac{\pi }{3}}d\varphi \ 
\Psi _{k,\ell,m,n}(r,\alpha,\theta ,\varphi )\Psi_{k^{\prime },\ell^{\prime },m^{\prime},n^{\prime }}(r,\alpha,\theta ,\varphi ) \nonumber\\
 & & =\delta_{k,k^{\prime }}\delta _{\ell,\ell^{\prime }} \delta _{m,m^{\prime }}  \delta _{n,n^{\prime }}
 N_{k,\ell,m,n} \ .
\label{3c.40.bis}
\end{eqnarray}%
Use is made, here,  of the orthogonality properties of 
 Gegenbauer, Jacobi and Laguerre polynomials \cite{Abramowitzbook}. 
The full expression of the  eigenenergies is  expressed by 
\begin{eqnarray}
E_{k,\ell,m,n} &\equiv &E_{k,\ell+2m+3n}=2\omega \left( 2k+1+\sqrt{\mu +(\ell+2m+3n+c+3a+3)^{2}  \ } \right) ,  \label{3c-42} \\
k &=&0,1,2,...,\qquad \ell=0,1,2,...\qquad m=0,1,2,...\qquad n=0,1,2,...,\;.  \nonumber
\end{eqnarray}

Let us now consider  the irregular solutions  corresponding to $1/2-a$.  We have to replace $a$ by $-a$ in all equations,
 from Eq.(\ref{3c-22})  to Eq.(\ref{3c-42}).
We recall that for $-1/2 < \lambda <3/2$, these irregular solutions are square integrable as mentioned before.
 For  Fermi  statistics,  a $\delta$ pathology  occurs in Eq.(\ref{comp})  for $a=1/2$ ($\lambda=0$).
Moreover,  the  requirement of self-adjointness of the Sturm-Liouville operator Eq.(\ref{3c-24}) 
imposes us  to ensure $b_n \neq 0$ ( i.e.  to discard the case $\lambda=0$) and $\beta >-1/4 \ (g > -3)$. 
Consequently, we consider values of $\lambda$ in $]-1/2,0[ \cup ]0,3/2[$.

Concerning the change $a \mapsto -a $ in   the function $\Theta_{m,n}(\theta)$,
Eqs.(\ref{3g-17},\ref{3gp-17}),  the self-adjointness of the Sturm-Liouville operator
imposes $B_n \ne 0,  \forall n$. Taking into account Eq.(\ref{3c-22-bis}), 
 we have clearly to look for $a < 1/2$, which is equivalent to $\lambda < 0$.
Then we search for which value of $a$, square integrable solutions
$\Theta_{m,n}(\theta)$   are obtained for every values of $n$. Since
\be
\frac{1}{2} +b_n=3 n + 2 -3 a \quad (a=\frac{1}{2} \sqrt{1+2 \lambda})  \ ,
\label{irr0}
\ee
we have
\be
(\forall n \geq 0) \qquad   \frac{1}{2}+b_n  \geq  2- 3 a  \ .
\label{irr1}
\ee
The function $\Theta_{m,n}(\theta)$, 
Eqs.(\ref{3g-17},\ref{3gp-17}), leads to square integrable solutions for every  $n$ if $a<5/6$.
This last inequality happens for  $\lambda <8/9$.
As far as the term $\beta/\cos^2 \theta$ is concerned,  irregular solutions are found when $c$ is changed to $-c$ in Eqs.(\ref{3g-17},\ref{thetap},\ref{3gp-17}).
These solutions are square integrable for $c < 1$  {\it i.e.}  for $\beta <3/4$ or $ g < 9$.
Consider the last angular  equation concerning the variable $\alpha$. The differential operator is self-adjoint
 provided  that 
\be
(\forall m \geq 0) \  (\forall n \geq 0) \qquad c_{m,n} = 2 m + 3 n +\frac{5}{2} \pm c  - 3 a  \geq 
\frac{5}{2}- 3 a \pm c  \ > 0 \ . 
\label{irr2}
\ee
So that the self-adjointness is ensured when 
\be
5 >  3 \sqrt{1+2 \lambda} \mp \sqrt{1+\frac{g}{3}} \ .
\ee
The square integrability condition for  the function $G_{\ell,m,n}$, Eq.(\ref{3c-30}),  requires 
%{\bf taking into account the Jacobian in Eq.(\ref{3c.40.bis})},
  $c_{m,n}+1 >0 \ \forall m \ \forall n$,
which yields
\be
\pm c - 3 a + \frac{7}{2} >0 \ .
\ee
This defines an acceptable domain of solutions in $\lambda,g$.

  As far as the radial equation is concerned, 
the constraint on $\mu + D_{\ell,m,n} >0$ is satisfied for $\mu >0$ otherwise  
it reads
\be
\vert 3 \pm c -3 a  \vert > \sqrt{ -\mu  \ } \ .
\ee
This condition defines a domain of acceptable values of $\mu$ depending on the values of 
$\lambda,g,$ within the interval  $\lambda \in ]-1/2,0[ \cup ]0,3/2[$.
  Under such conditions, the radial solutions, Eq.(\ref{3c-37}),  are square integrable because
 $\kappa_{\ell,m,n} >0$.

Note that the spectrum for irregular solutions has eigenvalues lower than the ones corresponding to the  regular solutions. They are given by
\begin{eqnarray}
E_{k,\ell,m,n}^{(<)} &\equiv &E_{k,\ell+2m+3n}^{(<)}=2\omega \left( 2k+1+\sqrt{\mu +(\ell+2m+3n \pm c-3a+3)^{2} \ } \right) ,  \label{3c-42m} \\
k &=&0,1,2,...,\qquad \ell=0,1,2,...\qquad m=0,1,2,...\qquad n=0,1,2,...,\;.  \nonumber
\end{eqnarray}

\subsection{The case of  Coulomb-type interaction}
In their work on solvable three-body problems in $D=1$, Khare and Badhuri \cite{KB} have considered
an attractive interaction of the form -const.$/\sqrt{\sum_{i <j} (x_i-x_j)^2}$. In spherical coordinates, such a term leads
to an attractive $1/r$ potential. %Replacing the harmonic field, it 
It can be used to replace the harmonic field in Hamiltonian like (\ref{3c-1}) and
produces Coulomb-like radial wave functions.
Following this idea we consider   the Hamiltonian
\begin{equation}
H=-\sum_{i=1}^{4} \frac{\partial ^{2}}{\partial x_{i}^{2}}%
-\frac{\eta}{\sqrt{ \sum_{i=1}^4 x_{i}^{2} \ } } +\lambda \sum_{i<j}^3 \frac{1}{%
(x_{i}-x_{j})^{2}}+\frac{g}{(x_1+x_2+x_3- 3  x_4)^2} +\frac{\mu }{\sum_{i=1}^{4}x_{i}^{2}},
\quad \eta >0  
\label{3c-c1}
\end{equation}%
The same coordinates transformations as Eq.(\ref{3c-16bis}) and  Eq.(\ref{3c-18bis})
are  applied to equation Eq.(\ref{3c-c1}), which becomes
\begin{eqnarray}
&&\left\{ -\frac{\partial ^{2}}{\partial r^{2}}-\frac{3}{r}\frac{\partial }{%
\partial r}-\frac{\eta}{r}+\frac{\mu }{r^{2}}  +  \frac{1}{r^{2}}\left[
 -\frac{\partial ^{2}}{\partial \alpha ^{2}}   -2 \cot \alpha \frac{\partial }{\partial \alpha } + 
\frac{1}{\sin ^{2}\alpha}  \left(  -\frac{\partial ^{2}}{\partial \theta ^{2}}-\cot \theta \frac{\partial }{\partial \theta }
 \right.\right. \right. \nonumber\\
 & & \left.\left.\left. + \frac{g}{12 \cos ^{2}\theta }+ \frac{1}{\sin ^{2}\theta }\left(- 
\frac{\partial ^{2}}{\partial \varphi ^{2}}+\frac{9\lambda }{2\sin
^{2}(3\varphi )}\right) \right) \right] \right\} \Psi(r,\alpha,\theta,\varphi)  =E\Psi(r,\alpha,\theta,\varphi)   \ . \label{3c-49b}
\end{eqnarray}%
Here, $\Psi (r,\alpha,\theta ,\varphi )$ are the eigensolutions associated to eigenenergy $E$.
Compared  to Eq.(\ref{3c-49}), the harmonic confinement has been replaced by 
a Coulomb-like field.

The resolution of   Eq.(\ref{3c-49b}) is achieved  in a similar manner as in the preceding section.
The solutions of the angular parts are identical. Only the radial part is different.
 Using the same factorization of the wave function as 
the one of Eq.(\ref{3c-6}), we obtain four decoupled differential equations.
The eigensolutions and eigenvalues  of the angular differential equations  are given
 respectively by Eqs.(\ref{3c-22},\ref{3c-22-bis}) for the variable $\varphi$,
by Eqs.(\ref{3g-17},\ref{3gp-17}) for the variable $\theta$ and 
 by Eqs.(\ref{3c-30},\ref{3c-31}) for the variable $\alpha$.

 The reduced radial equation reads 
\begin{equation}
\left( -\frac{d^{2}}{dr^{2}} -\frac{\eta}{r} +\frac{\mu +D_{\ell,m,n}-\frac{1}{4}}{%
r^{2}}-E_{k,\ell,m,n} \right) F_{k,\ell,m,n}(r)=0.  \label{+}
\end{equation}%
As in the case of the harmonic oscillator, the solutions for bound states ( negative energies) are well known.
The eigenfunctions are given by 
\cite{messiah,newton}
\begin{eqnarray}
F_{k,\ell,m,n}(r) & = &r^{\kappa _{\ell,m,n}+\frac{1}{2}}\exp \left(-  \tilde \eta \ r\right)L_{k}^{(2 \kappa _{\ell,m,n})}
(2 \tilde \eta r ) \nonumber\\
 & &  k=0,1,2...,  \quad \tilde \eta = \frac{\eta}{1 + 2 k + 2 \kappa _{\ell,m,n}}  \label{3c-37c} \ .
\end{eqnarray}
The associated eigenenergies take the form
\begin{equation}
E_{k,\ell,m,n}=-\frac{4 \eta^2}{(2k+\kappa _{\ell,m,n}+1)^2} ,\qquad k=0,1,2....  \label{3c-38c}
\end{equation}
with $\kappa _{\ell,m,n}$ given by Eq.(\ref{3c-33}).

In addition to the bound states, the equation Eq.(\ref{+}) admits scattering states
corresponding to  positive energies. They are  not studied in this paper.

\section{A five-body problem with  harmonic confinement.}

We consider the Hamiltonian
\begin{eqnarray}
H&=&\sum_{i=1}^{5}\left( -\frac{\partial ^{2}}{\partial x_{i}^{2}}%
+\omega ^{2}x_{i}^{2}\right) + \frac{\mu}{\sum_{i=1}^5 x_i^2} \nonumber\\
 & + &\lambda \sum_{i<j}^3 \frac{1}{%
(x_{i}-x_{j})^{2}}+ \frac{\kappa}{(x_4-x_5)^2} + \frac{g}{[2 (x_1+x_2+x_3)-  3 (x_4+x_5)]^2} 
\label{3ca-1}
\end{eqnarray}%
 The first contribution gives the energy of the five
independent particles in an harmonic field. Residual interactions are added in a following way.  
  The  first three particles,  with coordinates $x_1,x_2,x_3$,  constituting the first cluster, 
     interact pairwise  by a two-body inverse square potentials, of Calogero type \cite{Calo69}. 
     The remaining two-particles, whose coordinates are $x_4,x_5$, constituting the second cluster,
   interact also via a  two-body inverse square Calogero potential.
 The  centre of mass of both clusters interact  via the  translationally invariant  term 
 $g/(2 (x_1+x_2+x_3)-  3 (x_4+x_5))^2$.  
 An additional   five-body  potential, non translationally invariant, represented by the term 
$\mu/(x_1^2+x_2^2+x_3^2+x_4^2+x_5^2)$, is added to the whole Hamiltonian.

In order to solve this five-body problem,
let us introduce the following coordinates transformation 
\begin{eqnarray}
 & t &=\frac{1}{\sqrt{5}}(x_{1}+x_{2}+x_{3}+x_4+x_5),\quad u=\frac{1}{\sqrt{2}}(x_{1}-x_{2}),\quad v=\frac{1}{\sqrt{6}}(x_{1}+x_{2}-2x_{3}) \nonumber\\
   & z &=\frac{1}{\sqrt{2}}(x_{4}-x_{5}),
 \quad w=\frac{1}{\sqrt{30}} (2 (x_1+x_2+x_3)- 3 ( x_4+x_5)).
\label{3ca-16bis}
\end{eqnarray}

The transformed Hamiltonian reads:
\begin{eqnarray}
H &=&-\frac{\partial ^{2}}{\partial t^{2}}-\frac{\partial ^{2}}{\partial w^{2}}-\frac{\partial ^{2}}{\partial z^{2}} -\frac{%
\partial ^{2}}{\partial u^{2}}-\frac{\partial ^{2}}{\partial v^{2}}+\omega
^{2}(t^{2}+u^{2}+v^{2}+z^2+w^2)      \nonumber \\ 
& & +\frac{9\lambda (u^{2}+v^{2})^2}{%
2\left( u^{3}-3uv^{2}\right) ^{2}}  +\frac{\kappa}{2 z^2} +  \frac{g }{30 \ w^2}.  \label{3ca-48}
\end{eqnarray}

This Hamiltonian is not  separable in  $\{t,u,v,w,z\}$ variables.
As  before, we  introduce   hyperspherical coordinates : 
\begin{eqnarray}
t &=& r\cos \alpha,  \quad\quad\quad\quad w  = r\sin \alpha \cos \theta \nonumber\\
     0 & \leq &  r<\infty      \quad\quad\quad\quad\quad  0 \leq \alpha \leq \pi  \nonumber\\
z & = & r\sin \alpha \sin \theta \cos \beta, \quad u=r\sin \alpha \sin \theta \sin \beta  \sin \varphi,
\quad v=r\sin \alpha \sin \theta \sin \beta \cos \varphi ,\quad  \nonumber \\
  0 & \leq &  \theta \leq \pi, \quad\quad \quad\quad\quad\quad\quad  0\leq \beta \leq \pi, \quad \quad\quad\quad\quad\quad\quad\quad  0\leq \varphi \leq 2\pi .  \label{3ca-18bis}
\end{eqnarray}

The Schr\"{o}dinger equation is then written:
\begin{eqnarray}
&&\left\{ -\frac{\partial ^{2}}{\partial r^{2}}-\frac{4}{r}\frac{\partial }{%
\partial r}+\omega ^{2}r^{2}+\frac{\mu }{r^{2}}  +  \frac{1}{r^{2}}\left[
 -\frac{\partial ^{2}}{\partial \alpha ^{2}}   -3 \cot \alpha \frac{\partial }{\partial \alpha } + 
\frac{1}{\sin ^{2}\alpha}  \left(  -\frac{\partial ^{2}}{\partial \theta ^{2}}- 2 \cot \theta \frac{\partial }{\partial \theta }
 \right.\right. \right. \nonumber\\
 & & \left.\left.\left. + \frac{g}{30 \cos ^{2}\theta } + \frac{1}{\sin^2 \theta}  \left(   -\frac{\partial ^{2}}{\partial \beta ^{2}}-  \cot \beta \frac{\partial }{\partial \beta } + \frac{\kappa}{2 \cos^2 \beta} + \right.\right.\right.\right. \nonumber\\
 & &  \left.\left.\left.\left.  \frac{1}{\sin ^{2}\beta }\left(- 
\frac{\partial ^{2}}{\partial \varphi ^{2}}+\frac{9\lambda }{2\sin
^{2}(3\varphi )}\right) \right) \right) \right] \right\} \Psi(r,\alpha,\theta,\beta,\varphi)  =E\Psi(r,\alpha,\theta,\beta,\varphi)   \ .  \label{3ca-49}
\end{eqnarray}%
This Hamiltonian may be mapped to
 the problem of one particle in the five dimensional space with a non central potential of the form
\begin{equation}
V(r,\alpha,\theta ,\varphi )=f_{1}(r)+\frac{1}{r^{2}\sin ^{2}\alpha } \left[
f_{2}(\theta )+\frac{1}{\sin^2 \theta}  \left( f_{3}(\beta  ) +  \frac{f_{4}(\varphi )}{\sin ^{2}\beta } \right)  \right].
\end{equation}
As in the previous section, the problem becomes separable but in the five variables $\{r,\alpha,\theta,\beta,\varphi\}$.
To find the solution we factorize the wave function as follows :
\begin{equation}
\Psi_{k,\ell,j,m,n} (r,\alpha,\theta ,\beta,\varphi )=\frac{F_{k,\ell,j,m,n}(r)}{r^2} \frac{G_{\ell,j,m,n}(\alpha)}{\sin^{3/2} \alpha} \frac{\Theta_{j,m,n} (\theta )}{\sin\theta } \frac{H_{m,n}(\beta)}{\sqrt{\sin \beta}}
\Phi_n (\varphi ).  \label{3ca4-6} 
\end{equation}
Accordingly,  equation (\ref{3ca-49}) separates in  five   decoupled  differential equations:
\begin{equation}
\left( -\frac{d^{2}}{d\varphi ^{2}}+\frac{9\lambda }{2\sin^{2}(3\varphi )}%
\right) \Phi_n (\varphi )=B_n\Phi_n (\varphi ),\quad  \label{3ca-7}
\end{equation}

\begin{equation}
\left( -\frac{d^{2}}{d\beta^{2}}+\frac{(B_n-\frac{1}{4})}{\sin ^{2}\beta}%
+\frac{\kappa}{2  \cos^2\beta } \right) H_{m,n} (\beta )=C_{m,n} H_{m,n} (\beta ),\qquad  \label{3ca5-8}
\end{equation}%

\begin{equation}
\left( -\frac{d^{2}}{d\theta ^{2}}+\frac{(C_{m,n}-\frac{1}{4})}{\sin ^{2}\theta }%
+\frac{g}{30 \cos^2\theta} \right) \Theta_{j,m,n} (\theta )=D_{j,m,n} \Theta_{j,m,n} (\theta ),\qquad  \label{3ca-8}
\end{equation}%
\begin{equation}
\left( -\frac{d^{2}}{d\alpha ^{2}}+\frac{D_{j,m,n}-\frac{1}{4}}{\sin ^{2}\alpha }%
\right) G_{\ell,j,m,n} (\alpha )=A_{\ell,j,m,n} G_{\ell,j,m,n} (\alpha ),\qquad  \label{3ca-9}
\end{equation}%
and 
\begin{equation}
\left( -\frac{d^{2}}{dr^{2}}+\omega ^{2}r^{2}+\frac{\mu +A_{\ell,j,m,n}-\frac{1}{4}}{r^{2}%
}\right) F_{k,\ell,j,m,n}(r)=E_{k,\ell,j,m,n}\  F_{k,\ell,j,m,n}(r) \ .  \label{3ca-10}
\end{equation}

The regular solutions of equation (\ref{3ca-7})  on the interval $[0,\pi/3]$,  with Dirichlet conditions at the boundaries, are given  by the expressions (see above section)
\begin{equation}
\Phi _{n}(\varphi ) = (\sin 3\varphi )^{\frac{1}{2} + a}C_{n}^{(\frac{1}{2}
 + a)}(\cos 3\varphi ),     \quad 0\leq \varphi \leq \frac{\pi }{3},    
          \qquad n=0,1,2,...
\label{3ca-22} 
\end{equation}
and   associated to the eigenvalues 
\begin{equation}
b_{n}^{2}=B_{n} ,\quad b_{n}=\sqrt{B_{n}}=\left( 3n + 3 a+\frac{3}{2} \right), n=0,1,2,....  \quad  a =\frac{1}{2} \sqrt{1+2\lambda }  \label{3ca-21}
\end{equation}

The second angular equation for the polar angle $\beta$ reads : 
\begin{equation}
\left( -\frac{d^{2}}{d\beta ^{2}}+\frac{b_{n}^{2}-\frac{1}{4}}{\sin
^{2}\beta }+\frac{\kappa}{2 \cos ^2\beta} -C_{m,n}\right) H_{m,n} (\beta )=0 \ .  \label{3ca-24}
\end{equation}%
 We have $b_n  \ne  0$ so that the Hamiltonian of Eq.(\ref{3ca-24}) is a self-adjoint operator for $\kappa >-1/2$.

The regular solutions  of Eq.(\ref{3ca-24})  in $]0,\pi/2[$ with Dirichlet conditions are
 \begin{eqnarray}
H^{(+)} _{m,n}(\beta) &=&(\sin \beta )^{b_{n}+\frac{1}{2}}(\cos \beta )^{c+%
\frac{1}{2}}P_{m}^{(b_{n},c)}(\cos 2\beta ),\qquad  \label{3ca-17} \\ 
%ga-17
\quad 0 &\leq &\beta \leq \frac{\pi }{2},\quad m=0,1,2,...,\quad c=\frac{1}{%
2}\sqrt{1 + 2 \kappa \ } \  ,    
\label{thetapa}
\end{eqnarray}
 in terms of   the Jacobi polynomials. The solutions in the  interval $\frac{\pi }{2} < \beta <  \,\pi $ are 
\begin{eqnarray}
H^{(-)} _{m,n}(\beta &=&(\sin \beta
)^{b_{n}+\frac{1}{2}}(-\cos \beta)^{c+\frac{1}{2}}P_{m}^{(b_{n},c)}(\cos
2\beta ), \label{3gpa-17} \qquad \\
\frac{\pi}{2} & < &\beta  < \pi ,\quad m=0,1,2,...\,.  \nonumber
\end{eqnarray}
The index  $+$ (resp.$-$) means that the 
$z$-coordinate is positive. (resp. negative).
Note that in the interval $]\pi/2,\pi[$, we have $-\cos \beta > 0 $ and $\sin \beta > 0$, so that 
the real power of these positive values are defined.

In both cases, the eigenvalues  $C_{m,n}$ of Eq.(\ref{3ca-24}) are given by 
\begin{equation}
C_{m,n}=c_{m,n}^2, \qquad  c_{m,n}=(2 m +b_{n}+c+1) ,\quad m=0,1,2,...\,.\quad
\label{Cmna}
\end{equation}

The third angular equation for the polar angle $\theta$ can be written as
\begin{equation}
\left( -\frac{d^{2}}{d\theta ^{2}}+\frac{c_{m,n}^2-\frac{1}{4}}{\sin
^{2}\theta }+\frac{g}{30 \cos ^2\theta} -D_{j,m,n}\right) \Theta_{j,m,n} (\theta )=0,  \label{3cap-24}
\end{equation}%

Since  $c_{m,n} \ne  0$, (see Eqs.(\ref{3ca-21},\ref{Cmna})) the Hamiltonian of Eq.(\ref{3cap-24}) is a self-adjoint operator provided that $g  > -15/2$.

The regular solutions  Eq.(\ref{3cap-24}) in $]0,\pi/2[$ with Dirichlet conditions read
 \begin{eqnarray}
\Theta^{(+)} _{j,m,n}(\theta ) &=&(\sin \theta )^{c_{m,n}+\frac{1}{2}}(\cos \theta )^{d+%
\frac{1}{2}}P_{m}^{(c_{m,n},d)}(\cos 2\theta ),\qquad  \label{3ga-17} \\
\quad 0 &\leq &\theta \leq \frac{\pi }{2},\quad m=0,1,2,...,\quad d=\frac{1}{2} \sqrt{1+\frac{2 g}{15} }     \  .      
\label{thetapb}
\end{eqnarray}

 The solutions in the  interval $\frac{\pi }{2} < \theta <  \,\pi $ are 
\begin{eqnarray}
\Theta^{(-)} _{j,m,n}(\theta) &=&(\sin \theta
)^{c_{m,n}+\frac{1}{2}}(-\cos \theta )^{d+\frac{1}{2}}P_{j}^{(c_{m,n},d)}(\cos
2\theta ), \label{3gpaa-17} \qquad \\
\frac{\pi}{2} & < &\theta < \pi ,\quad m=0,1,2,...\,.  \nonumber
\end{eqnarray}
The index $+$ (resp.$-$) means  the $w$-coordinate to be positive.  (resp. negative).
Note that in the interval $]\pi/2,\pi[,$ we have $-\cos \theta > 0 $ and $\sin \theta > 0,$ so that 
the real power of these positive values are defined.

In both cases, the eigenvalues  $D_{j,m,n}$ of Eq.(\ref{3cap-24}) are given by 
\begin{equation}
D_{j,m,n}=d_{j,m,n}^2 \quad d_{j,m,n}=(2 j +c_{m,n}+d+1), \quad j=0,1,2,...\,.\quad
\label{Dmna}
\end{equation}

The regular eigensolutions and corresponding eigenvalues of  equation (\ref{3ca-9}) in the interval $]0,\pi[$
read, respectively, 
\begin{eqnarray}
G _{\ell,j,m,n}(\alpha ) &=&(\sin \alpha )^{d_{j,m,n}+\frac{1}{2}}C_{\ell}^{(d_{j,m,n}+%
\frac{1}{2})}(\cos \alpha ),\qquad \ell=0,1,2,...,\quad  \label{3ca-30} \\
A_{\ell,j,m,n} &=&a_{\ell,j,m,n}^2, \quad a_{\ell,j,m,n}=
\left( \ell+d_{j,m,n}+\frac{1}{2} \right) \qquad  \,\, \ell=0,1,2,...\ .
\label{3ca-31}
\end{eqnarray}

Note that the choice  $d_{j,m,n} >0$ implies  for every value of  $\{j,m,n\}$ the function
$G_{\ell,j,m,n}(\alpha)$  to vanish at the boundaries  of the interval  $\left ]0,\pi \right[.$

Finally, the reduced radial equation reads 
\begin{equation}
\left( -\frac{d^{2}}{dr^{2}}+\omega ^{2}r^{2}+\frac{\mu +A_{\ell,j,m,n}-\frac{1}{4}}{%
r^{2}}-E_{k,\ell,j,m,n} \right) F_{k,\ell,j,m,n}(r)=0.  \label{3ca-32}
\end{equation}%

Taking into account the definition of $A_{\ell,j,m,n}$,  Eq.(\ref{3ca-31}), we have
\begin{equation}
\mu +A_{\ell,j,m,n}=\mu +\left(\ell+2 j + 2 m + 3 n + c + d +  3  a +  4 \right)^{2}>0  \quad
\forall n\geq 0, \forall m\geq 0,  \forall \ell \geq 0 \ ,  \label{3ca-32bispp}
\end{equation}
for every positive $\mu$. 
The quantity $\mu +A_{\ell,j,m,n}$ is minimal for  $n=0,j=0,m=0,\ell=0$ and $a=0$ ( we recall that $a \geq 0$, 
see (\ref{3ca-21}) and that $c \geq 0$, see (\ref{thetapa})). 
It put constraint on negative values of  $\mu$,  namely $- 16 <  \mu \leq  0. $

%\begin{equation}
% - 16 <  \mu \leq  0 
%\end{equation}
 We introduce the auxiliary parameter  $\kappa _{\ell,j,m,n}$ defined by
\begin{equation}
\kappa _{\ell,j,m,n}^{2}=\mu +A_{\ell,j,m,n} \ , \qquad \kappa _{\ell,j,m,n}=
\sqrt{\mu +A_{\ell,j,m,n} } \ .
\label{3ca-33}
\end{equation}
 The solution of the radial equation (\ref{3ca-32})  is

\begin{equation}
F_{k,\ell,j,m,n}(r)=r^{\kappa _{\ell,jm,n}+\frac{1}{2}}\exp \left(-\frac{\omega r^{2}}{2}
\right)L_{k}^{(\kappa _{\ell,j,m,n})}(\omega r^{2}),\qquad k=0,1,2... \ .  \label{3ca37}
\end{equation}
in terms of the generalized Laguerre polynomials. The associated eigenenergies are given by
\begin{equation}
E_{k,\ell,j,m,n}=2\omega (2k+\kappa _{\ell,j,m,n}+1),\qquad k=0,1,2....  \label{3ca-38}
\end{equation}

Collecting all pieces, we conclude  the physically acceptable regular solutions of 
the Schr\"odinger equation  (\ref{3ca-49}) to be given in  a compact form ( as explained in the previous section) by 
\begin{eqnarray}
\Psi _{k,\ell,j,m,n}(r,\alpha,\theta , \beta, \varphi ) &=&r^{\sqrt{\mu + \left(\ell+2j+2m+3n+c+d+3a+4 \right)^{2} \ }-3/2}e^{-\frac{\omega r^{2}}{2}}L_{k}^{\left(\sqrt{\mu + \left(\ell+2j+2m+3n+c+d+3a+3 \right)^{2}    \ }\right) }(\omega r^{2})  \nonumber \\
& &\times (\sin \alpha )^{2j+2m+3n+c+d+3a+\frac{5}{2}}C_{\ell}^{^{\left(2j+2m+3n+c+d+4 \right)
}}(\cos \alpha )  \nonumber \\
& &\times (\sin \theta )^{2m+3n+3a+c+2} (\epsilon_2 \cos \theta)^{d+\frac{1}{2}} P_{j}^{\left( 2m+3n+c+3a+\frac{5}{2},d \right)
}(\cos 2 \theta )  \nonumber \\
 & & \times (\sin \beta )^{3n+3a+\frac{3}{2}} (\epsilon_1 \cos \beta)^{c+\frac{1}{2}} P_{m}^{\left( 3n+3a+\frac{3}{2},d \right)}(\cos 2 \beta )  \nonumber \\
& & \times sgn(\sin 3\varphi)^{[(1-\epsilon_3)/2]}   \vert \sin 3\varphi \vert^{a+\frac{1}{2}}C_{n}^{\left( a+\frac{1}{2}\right)}(\cos 3\varphi ),\,     \label{compa} \\
 k & = & 0,1,2,...,\quad \ell=0,1,2,...,\quad j=0,1,2,.., \quad m=0,1,2,..,\quad n=0,1,2,...,   \nonumber \\
 & & 0  \leq  \varphi \leq \frac{\pi }{3}\ , \quad    \frac{1-\epsilon_1}{2} \frac{\pi}{2}  \leq \beta \leq \frac{3-\epsilon_1}{2} \frac{\pi }{2}, \epsilon_1=\pm 1 \nonumber\\ 
 & &  \frac{1-\epsilon_2}{2} \frac{\pi}{2}  \leq   \theta \leq 
\frac{3-\epsilon_2}{2} \frac{\pi }{2} , \epsilon_2=\pm 1, \qquad\quad 0 \leq \alpha \leq \pi  \nonumber\\
& &  \quad a=\frac{1}{2}\sqrt{
1+2\lambda } \ ,\quad c=\frac{1}{2}\sqrt{1 + 2 \kappa \ } \quad  d=\frac{1}{2} \sqrt{1 + \frac{2g}{15} \ } \ . \quad \quad \nonumber    
\end{eqnarray}

Here, $\epsilon_3=1$ for bosons and $-1$ for fermions in equation Eq.(\ref{compa}).
 For the Bose statistics,  the extension  (\ref{compa}) is possible   provided that    no $\delta$ distribution 
appears when the second derivative of the wave function 
is applied  at the boundaries  between two adjacent sectors.
For  $\varphi=\pi/3$ and $n=0$, a $\delta$ distribution occurs for 
$a=1/2$ implying  $\lambda=0$, due to the term $\vert \sin 3 \varphi \vert$ in (\ref{compa}). 
Also a $\delta$ distribution occurs for $d=1/2$ i.e. $g=0$ at  $\theta=\pi/2$.
Consequently,   the symmetrical solutions 
 for the pure harmonic oscillator ($\lambda=\mu=g=0$) are not recovered.

The normalization constants $N_{k,\ell,j,m,n}$ can be  calculated from
\begin{eqnarray}
&& \int_{0}^{+\infty }r^{4}dr\int_{0}^{\pi }\sin^3 \alpha \  d\alpha 
\int_{(1-\epsilon_2) \pi/4}^{(3-\epsilon_2) \pi/4} \sin^2  \theta \  \ d\theta
\int_{(1-\epsilon_1) \pi/4}^{(3-\epsilon_1) \pi/4} \sin \beta \ d\beta \nonumber\\
 & &  \times \int_0^{\frac{\pi }{3}}d\varphi 
\Psi _{k,\ell,j,m,n}(r,\alpha,\theta ,\beta,\varphi )\Psi_{k^{\prime },\ell^{\prime },j^{\prime },m^{\prime },n^{\prime }}
(r,\alpha,\theta ,\beta,\varphi )  =\delta_{k,k^{\prime }}\delta _{\ell,\ell^{\prime }}\delta _{j,j^{\prime }}\delta _{m,m^{\prime }}  \delta _{n,n^{\prime }} N_{k,\ell,j,m,n} \ .
\label{3ca.40.bis}
\end{eqnarray}%
As above, use is made  of the orthogonality properties of 
 Gegenbauer, Jacobi and Laguerre polynomials.

The full expression of the  eigenenergies is  expressed by 
\begin{eqnarray}
E_{k,\ell,j,m,n} & \equiv &E_{k,\ell+ 2j+2m+3n}=2\omega \left( 2k+1+\sqrt{\mu+(\ell+2j+2m+3n+c+d+3a+4)^{2}  \ } \right)  \nonumber\\
k &=&0,1,2,...,\quad \ell=0,1,2,... \quad j=0,1,2, \quad m=0,1,2,...\quad n=0,1,2,... \ .  \label{3ca-42}
\end{eqnarray}

\section{A six-body  problem with  harmonic confinement.}

We consider the Hamiltonian
\begin{eqnarray}
H &= &\sum_{i=1}^{6}\left( -\frac{\partial ^{2}}{\partial x_{i}^{2}}%
+\omega ^{2}x_{i}^{2}\right) +\lambda_1 \sum_{i<j}^3 \frac{1}{(x_{i}-x_{j})^{2}}+
\lambda_2 \sum_{i<j=4}^6 \frac{1}{(x_{i}-x_{j})^{2}} \nonumber\\
  &+ & \frac{g}{(x_1+x_2+x_3-  x_4-x_5-x_6)^2} + \frac{\mu}{ \sum_{i=1}^{6} x_i^2} 
\label{3cb-1}
\end{eqnarray}

The first contribution gives the energy of the six
independent particles in an harmonic field. 
  The  first three particles, whose coordinates are $x_1,x_2,x_3$, constitute the first cluster  and the remaining three particles, whose coordinates are $x_4,x_5,x_6$, constitute the second cluster. 
In each cluster, the particles     interact pairwise  by a two-body inverse square potentials. 
The center  of mass  of both  clusters interact via the six-body     translationally invariant  
inverse square potential, namely $g/(x_1+x_2+x_3-  x_4-x_5-x_6)^2.$
A non-translationaly invariant six-body potential with coupling constant $\mu$ is added.

In order to solve this six-body problem,
let us introduce the following coordinates transformation 
\begin{eqnarray}
 & t &=\frac{1}{\sqrt{6}}(x_{1}+x_{2}+x_{3}+x_4+x_5+x_6),\quad u_1=\frac{1}{\sqrt{2}}(x_{1}-x_{2}),\quad v_1=\frac{1}{\sqrt{6}}(x_{1}+x_{2}-2x_{3}) \nonumber\\
  u_2 & = &\frac{1}{\sqrt{2}}(x_{4}-x_{5}), v_2=\frac{1}{\sqrt{6}}(x_{4}+x_{5}-2 x_{6}),
   \quad w=\frac{1}{\sqrt{6}} (x_{1}+x_{2}+x_{3}-x_4-x_5-x_6) 
\label{3cb-16bis}
\end{eqnarray}

The transformed Hamiltonian reads:
\begin{eqnarray}
H &=&-\frac{\partial ^{2}}{\partial t^{2}}-\frac{\partial ^{2}}{\partial w^{2}}-\frac{\partial ^{2}}{\partial u_1^{2}} -\frac{\partial ^{2}}{\partial u_2^{2}}-\frac{\partial ^{2}}{\partial v_1^{2}} -\frac{\partial ^{2}}{\partial v_2^{2}}
+\omega^{2}(t^{2}+w^{2}+u_1^2+u_2^2+v_1^2+v_2^2)      \nonumber \\ 
& &+ \frac{\mu}{t^{2}+w^{2}+u_1^2+u_2^2+v_1^2+v_2^2  } \ + \frac{9\lambda_1 (u_1^{2}+v_1^{2})^2}{2\left( u_1^{3}-3u_1v_1^{2}\right) ^{2}}  
 +\frac{9\lambda_2 (u_2^{2}+v_2^{2})^2}{2\left( u_2^{3}-3u_2v_2^{2}\right) ^{2}}  +
 \ \frac{g }{6 \ w^2}.  \label{3cb-48}
\end{eqnarray}
This Hamiltonian is not  separable in  $\{t,u_1,v_1,u_2,v_2,w\}$ variables.
We first introduce the following coordinates transformation : 
\begin{eqnarray}
 u_1 &= &r_1  \sin \varphi_1,\quad v_1=r_1 \cos \varphi_1 ,\quad0 \leq r_1<\infty, \quad 0\leq
\varphi_1 \leq 2\pi \nonumber\\
 u_2 &= &r_2 \sin \varphi_2,\quad v_2=r_2 \cos \varphi_2 ,\quad0 \leq r_2<\infty, \quad 0\leq
\varphi_2 \leq 2\pi 
\label{3cb-18bis}
\end{eqnarray}
The Schr\"{o}dinger equation is then written:
\begin{eqnarray}
&&\left\{-\frac{\partial ^{2}}{\partial t^{2}}-\frac{\partial ^{2}}{\partial w^{2}} -\frac{\partial ^{2}}{\partial r_1^{2}}-\frac{1}{r_1}\frac{\partial }{\partial r_1} -\frac{\partial ^{2}}{\partial r_2^{2}}-\frac{1}{r_2}\frac{\partial }{\partial r_2}+ \frac{g }{6 w^2} +  \omega ^{2}(r_1^{2}+r_2^2+t^{2}+w^2) \right. \nonumber\\
 & & \left.  +\frac{\mu}{r_1^{2}+r_2^2+t^{2}+w^2}
  +  \frac{1}{r_1^{2}} \left[-\frac{\partial ^{2}}{\partial \varphi_1^{2}}+\frac{9\lambda_1 }{2\sin
^{2}(3\varphi_1 )} \right]    \right. \nonumber\\
& & \left. + \frac{1}{r_2^{2}} \left[-\frac{\partial ^{2}}{\partial \varphi_2^{2}}+\frac{9\lambda_2 }{2\sin^{2}(3\varphi_2 )} \right]   \right\} \Psi(t,w,r_1,r_2,\varphi_1,\varphi_2)  =E\Psi(t,w,r_1,r_2,\varphi_1,\varphi_2)   \ .  \label{3cb-49}
\end{eqnarray}%
The potential involved in the equation (\ref{3cb-49})
\begin{eqnarray}
V(t,w,r_1,r_2,\varphi_1,\varphi_2) &= & \omega ^{2}(r_1^{2}+r_2^2+t^{2}+w^2)+
\frac{\mu}{r_1^{2}+r_2^2+t^{2}+w^2}+\frac{g }{6 w^2} \nonumber\\
 & +& \frac{1}{r_1^{2}}\  \frac{9\lambda_1 }{2\sin^{2}(3\varphi_1 )}   + \frac{1}{r_2^{2}} \  \frac{9\lambda_2 }{2\sin^{2}(3\varphi_2 )}
\label{potc}
\end{eqnarray}
 has the general form 
\begin{equation}
V(t,w,r_1,r_2,\varphi_1,\varphi_2)=f(t,w,r_1,r_2) + \frac{f_1(\varphi_1)}{r_1^2}+  \frac{f_2(\varphi_2)}{r_2^2} \ .
\end{equation}
This suggests the wave function to be factorized as follows
\begin{equation}
\Psi(t,w,r_1,r_2,\varphi_1,\varphi_2)  =\tilde \Psi(t,w,r_1,r_2) \prod_{j=1}^2  \Phi_j(\varphi_j )  \ .
\end{equation}
The equation (\ref{3cb-49})  will be solved  in two steps.
Firstly we solve
\begin{equation}
\left( -\frac{d^{2}}{d\varphi_1 ^{2}}+\frac{9\lambda_1 }{2\sin^{2}(3\varphi_1 )}%
\right) \Phi_{n_1}^{(1)} (\varphi_1 )=B_{n_1}^{(1)} \Phi_{n_1}^{(1)} (\varphi_1 ),\quad  \label{3cb1-7}
\end{equation}
and
\begin{equation}
\left( -\frac{d^{2}}{d\varphi_2 ^{2}}+\frac{9\lambda_2 }{2\sin^{2}(3\varphi_2 )}%
\right) \Phi_{n_2}^{(2)} (\varphi_2 )=B_{n_2}^{(2)} \Phi_{n_2}^{(2)} (\varphi_2 ),\quad  \label{3cb2-7}
\end{equation}
on the interval $[0,\pi/3]$,  with Dirichlet conditions at the boundaries.
The $B_{n_j}^{(j)}, j=1,2$  are the quantified eigenvalues of  the equations (\ref{3cb1-7},\ref{3cb2-7}), respectively  given by 
\begin{equation}
B_{n_j}^{(j)} = (b_{n_j}^{(j)})^2 \qquad b_{n_j}^{(j)}=3\left( n_j  + \frac{1}{2} + a_j \right),\qquad n_j=0,1,2,.... \quad j=1,2  \ . \label{3cb-22-bis}
\end{equation}
The associated eigensolutions are given in terms of the Gegenbauer polynomials $C_n^{(q)}$ 
\begin{equation}
\Phi _{n_j}^{(j)}(\varphi_j ) = (\sin 3\varphi_j )^{\frac{1}{2}+ a_j}C_{n_j}^{(\frac{1}{2}+
a_j)}(\cos 3\varphi_j ),     \quad 0\leq \varphi \leq \frac{\pi }{3},    
          \qquad n_j=0,1,2,.., \quad a_j =\frac{1}{2} \sqrt{1+2\lambda_j } \ .
\label{3cb-22} 
\end{equation}

The second step consists in the resolution of the  following Schr\"{o}dinger equation 

\begin{eqnarray}
&&\left\{-\frac{\partial ^{2}}{\partial t^{2}}-\frac{\partial ^{2}}{\partial w^{2}} -\frac{\partial ^{2}}{\partial r_1^{2}}-\frac{1}{r_1}\frac{\partial }{\partial r_1} -\frac{\partial ^{2}}{\partial r_2^{2}}-\frac{1}{r_2}\frac{\partial }{\partial r_2}+ \frac{g }{6 w^2} \right.  \nonumber\\
 &  & \left.+\omega ^{2}(r_1^{2}+r_2^2+t^{2}+w^2) +  \frac{\mu}{r_1^{2}+r_2^2+t^{2}+w^2} \right. \nonumber\\
 & & \left.  + \frac{B_{n_1}^{(1)}}{r_1^{2}} + \frac{B_{n_2}^{(2)}}{r_2^{2}}
 \right\} \tilde \Psi_{n_1,n_2}(t,w,r_1,r_2) =E_{n_1,n_2} \tilde \Psi_{n_1,n_2}(t,w,r_1,r_2) .  \label{3cbb-49}
\end{eqnarray}%

This  Hamiltonian   is not  separable in  $\{t,w,r_1,r_2\}$ variables.
We then introduce the following  hyperspherical coordinates : 
\begin{eqnarray}
t &=&r\cos \alpha, \quad w=r\sin \alpha \cos \theta, \quad r_1=r\sin \alpha \sin \theta \sin \beta
\quad r_2=r\sin \alpha \sin \theta \cos \beta  \nonumber \\
0 &\leq &r<\infty ,\quad \quad  0\leq \alpha \leq \pi ,\quad\quad\quad\quad   0\leq \theta \leq \pi,\quad \quad\quad\quad\quad 0\leq \beta  \leq \pi/2 .  \label{3cb-18biss}
\end{eqnarray}
Since $r_1,r_2$ are positive we have $\beta \in [0,\pi/2]$.

The Schr\"{o}dinger equation (\ref{3cbb-49}) becomes :
\begin{eqnarray}
& &\left\{ -\frac{\partial ^{2}}{\partial r^{2}}-\frac{5}{r}\frac{\partial }{%
\partial r}+\omega ^{2}r^{2}+\frac{\mu }{r^{2}}  +  \frac{1}{r^{2}}\left[
 -\frac{\partial ^{2}}{\partial \alpha ^{2}}   -4 \cot \alpha \frac{\partial }{\partial \alpha } \right.\right. \nonumber\\
  & &\left.\left.  + \frac{1}{\sin ^{2}\alpha}  \left(  -\frac{\partial ^{2}}{\partial \theta ^{2}}- 3 \cot \theta \frac{\partial }{\partial \theta } 
 + \frac{g}{6 \cos ^{2}\theta } + \frac{1}{\sin^2 \theta}  \left(   -\frac{\partial ^{2}}{\partial \beta ^{2}}-  2 \cot 2 \beta \frac{\partial }{\partial \beta }  \right.\right.\right.\right. \nonumber\\
 & &  \left.\left.\left.\left.  \frac{B_{n_1}^{(1)}}{\sin ^{2}\beta } + \frac{B_{n_2}^{(2)}}{\cos^2 \beta} 
 \right) \right) \right] \right\} \tilde \Psi_{n_1,n_2}(r,\alpha,\theta,\beta) 
 =E_{n_1,n_2} \tilde \Psi_{n_1,n_2}(r,\alpha,\theta,\beta)    .  \label{3cbp-49}
\end{eqnarray}%

This Hamiltonian may be mapped to
 the problem of one particle in the four dimensional space with a non central potential of the form
\begin{equation}
V(r,\alpha,\theta ,\beta)=f_{1}(r)+\frac{1}{r^{2}\sin ^{2}\alpha } \left[
f_{2}(\theta )+\frac{1}{\sin^2 \theta}   f_{3}(\beta  ) \right].
\end{equation}
As in the second section, the problem becomes separable in the four variables $\{r,\alpha,\theta,\beta\}$.
To find the solution we factorize the wave function as follows :
\begin{equation}
 \tilde \Psi_{n_1,n_2}(r,\alpha,\theta,\beta)=\frac{F_{k,\ell,j,m,n_1,n_2}(r)}{r^{5/2}} \frac{G_{\ell,j,m,n_1,n_2}(\alpha)}{\sin^2 \alpha} \frac{\Theta_{j,m,n_1,n_2} (\theta )}{\sin\theta^{3/2} } \frac{H_{m,n_1,n_2} (\beta)}{\sqrt{\sin 2 \beta}} \ . \label{3cb4-6} 
\end{equation}
Accordingly,  equation (\ref{3cbp-49}) separates in  four  decoupled  differential equations:

\begin{equation}
\left( -\frac{d^{2}}{d\beta^{2}}+\frac{(b_{n_1}^{(1)})^2-1/4}{\sin ^{2}\beta } + \frac{(b_{n_2}^{(2)})^2-1/4}{\cos^2 \beta} 
  \right) H_{m,n_1,n_2} (\beta )=  C_{m,n_1,n_2}  H_{m,n_1,n_2} (\beta ),\qquad  \label{3cb5-8}
\end{equation}%
in terms of the $b_{n_j}^{(j)}$ of Eq.(\ref{3cb-22-bis}).
\begin{equation}
\left( -\frac{d^{2}}{d\theta ^{2}}+\frac{(C_{m,n_1,n_2}-\frac{1}{4})}{\sin ^{2}\theta }%
+\frac{g}{6 \cos^2\theta} \right) \Theta_{j,m,n_1,n_2} (\theta )=D_{j,m,n_1,n_2} \Theta_{j,m,n_1,n_2} (\theta ),\qquad  \label{3cb-8}
\end{equation}%
\begin{equation}
\left( -\frac{d^{2}}{d\alpha ^{2}}+\frac{D_{j,m,n_1,n_2}-\frac{1}{4}}{\sin ^{2}\alpha }%
\right) G_{\ell,j,m,n_1,n_2} (\alpha )=A_{\ell,j,m,n_1,n_2} G_{\ell,j,m,n_1,n_2} (\alpha ),\qquad  \label{3cb-9}
\end{equation}%
and 
\begin{equation}
\left( -\frac{d^{2}}{dr^{2}}+\omega ^{2}r^{2}+\frac{\mu +A_{\ell,j,m,n_1,n_2
}-\frac{1}{4}}{r^{2}%
}\right) F_{k,\ell,j,m,n_1,n_2}(r)=E_{k,\ell,j,m,n_1,n_2}\  F_{k,\ell,j,m,n_1,n_2}(r) \ .  \label{3cb-10}
\end{equation}

The regular solutions of Eq.(\ref{3cb5-8}),  for $\beta \in [0,\pi/2]$ (remind that $r_1,r_2 \geq 0$), with Dirichlet conditions
at the boundaries read  
 
\begin{eqnarray}
H _{m,n_1,n_2}(\beta ) &=&(\sin \beta )^{b_{n_1}^{(1)}+\frac{1}{2}} \ (\cos \beta )^{b_{n_2}^{(2)}+
\frac{1}{2}} \ P_{m}^{(b_{n_1}^{(1)},b_{n_2}^{(2)})}(\cos 2\beta ),\qquad  \label{3gb-17} \\
\quad 0 &\leq &\beta  \leq \frac{\pi }{2},\quad m=0,1,2,...,\
\label{beta6}
\end{eqnarray}
%abelT2014
 in terms of the Jacobi polynomials $P_{m}^{( b_{n_1}^{(1)},b_{n_2}^{(2)}  )}$.
The associated eigenvalues  $C_{m,n_1,n_2}$ are given by 
\begin{equation}
C_{m,n_1,n_2}=c_{m,n_1,n_2}^2, \qquad c_{m,n_1,n_2}=(2 m +b_{n_1}^{(1)}+ b_{n_2}^{(2)}+1),\quad m=0,1,2,...\,.\quad
\label{Cmnb}
\end{equation}
Since $b_{n_j}^{(j)} \ne  0,  j=1,2 $  
the Hamiltonian of Eq.(\ref{3cb5-8}) is a self-adjoint operator.

The second angular equation for the polar angle $\theta$ reads : 
\begin{equation}
\left( -\frac{d^{2}}{d\theta ^{2}}+ \frac{c_{m,n_1,n_2}^{2}-1/4}{\sin ^{2}\theta} + \frac{d^2-1/4}{\cos^2 \theta}  -D_{j,m,n_1,n_2}\right) \Theta_{j,m,n_1,n_2} (\theta )= 0,  \label{3cb-24}
\end{equation}%
with $c_{m,n_1,n_2}$ given by Eq.(\ref{Cmnb}) and $d=\sqrt{g/6+1/4}$ with the condition that $g \geq -3/2$.
Since  $c_{m,n_1,n_2} \ne  0$, the Hamiltonian of Eq.(\ref{3cb-24}) is a self-adjoint operator if $d \ne 0$ i.e. $ g \ne -3/2$.

The regular solutions  of Eq.(\ref{3cb-24})  in $]0,\pi/2[$ with Dirichlet conditions
 read, in terms of  the Jacobi polynomials,
 \begin{eqnarray}
\Theta^{(+)} _{j,m,n_1,n_2}(\theta ) &=&(\sin \theta )^{c_{m,n_1,n_2}+\frac{1}{2}}(\cos \theta )^{d+%
\frac{1}{2}}P_{j}^{(c_{m,n_1,n_2},d)}(\cos 2\theta ),\qquad  \label{3gbb-17} \\
\quad 0 &\leq &\theta \leq \frac{\pi }{2},\quad j=0,1,2,...,\quad d=\frac{1}{2} \sqrt{1+\frac{2 g}{3} }     \  .      
\label{thetapab}
\end{eqnarray}

The index $+$ means that the 
$w$-coordinate is positive. The solutions in the  interval $\frac{\pi }{2} < \theta <  \,\pi $ are 
\begin{eqnarray}
\Theta^{(-)} _{j,m,n_1,n_2}(\theta) &=&(\sin \theta
)^{c_{m,n_1,n_2}+\frac{1}{2}}(-\cos \theta )^{d+\frac{1}{2}}P_{j}^{(c_{m,n_1,n_2},d)}(\cos
2\theta ), \label{3gpb-17} \qquad \\
\frac{\pi}{2} & < &\theta < \pi ,\quad j=0,1,2,...\,.  \nonumber
\end{eqnarray}
The index $-$ means the $w$-coordinate to be  negative.
Note that in the interval $]\pi/2,\pi[$ we have $-\cos \theta > 0 $ and $\sin \theta > 0$ so that 
the real power of these positive values are defined.

In both cases, the eigenvalues  $D_{j,m,n_1,n_2}$ of Eq.(\ref{3cb-24}) are given by 
\begin{equation}
D_{j,m,n_1,n_2}=d_{j,m,n_1,n_2}^2, \quad d_{j,m,n_1,n_2}=(2 j +c_{m,n_1,n_2}+d+1), \quad j=0,1,2,...\,.\quad
\label{Djmnb}
\end{equation}
%abdelT2014

The regular eigensolutions and corresponding eigenvalues in the interval $]0,\pi[$ of the third    equation (\ref{3cb-9})
read, respectively, 

\begin{eqnarray}
G _{\ell,j,m,n_1,n_2}(\alpha ) &=&(\sin \alpha )^{d_{j,m,n_1,n_2}+\frac{1}{2}} \ C_{\ell}^{(d_{j,m,n_1,n_2}+%
\frac{1}{2})}(\cos \alpha ),\qquad \ell=0,1,2,...,\quad  \label{3cb-30} \\
A_{\ell,j,m,n_1,n_2} &=&a_{\ell,j,m,n_1,n_2}^2,  \quad a_{\ell,j,m,n_1,n_2}=
\left( \ell+d_{j,m,n_1,n_2}+\frac{1}{2} \right) ,\quad \,\, \ell=0,1,2,...\ .
\label{3cb-31}
\end{eqnarray}
%abdelG2014

Note that the choice  $d_{j,m,n_1,n_2} >0$ implies  for every value of  $\{j,m,n_1,n_2\}$ the function
$G_{\ell,j,m,n_1,n_2}(\alpha)$  to vanish at the boundaries  of the interval  $\left ]0,\pi \right[.$

Finally, the reduced radial equation reads 
\begin{equation}
\left( -\frac{d^{2}}{dr^{2}}+\omega ^{2}r^{2}+\frac{\mu +A_{\ell,j,m,n_1,n_2}-\frac{1}{4} }{%
r^{2}}-E_{k,\ell,j,m,n_1,n_2} \right) F_{k,\ell,j,m,n_1,n_2}(r)=0.  \label{3cb-32}
\end{equation}%

Taking into account the definition of $A_{\ell,j,m,n_1,n_2}$,  Eq.(\ref{3cb-31}),  the constraint
\begin{eqnarray}
\mu +A_{\ell,j,m,n_1,n_2}& = &\mu +\left(\ell+2 j + 2 m + 3 n_1 + 3 n_2 + d + 3  a_1 + 3 a_2 + \frac{11}{2} 
  \right)^{2} > 0  \nonumber\\
&  &  \forall n_1\geq 0,\quad \forall n_2\geq 0,\quad \forall m\geq 0, \quad \forall j \geq 0, \quad \forall \ell \geq 0,  \label{3cb-32bisp}
\end{eqnarray}
happens, for every positive value of $\mu$. Since  $\mu +A_{\ell,j,m,n}$ is minimal for  $n=0,j=0,m=0,\ell=0$ and $a_1=a_2=d=0,$ it put constraint on  negative values of $\mu$, which has to satisfy  
\begin{equation}
 - \left( \frac{\sqrt{11}}{2}\right)^2 <  \mu \leq  0 
\end{equation}

 We introduce the auxiliary parameter  $\kappa _{\ell,j,m,n_1,n_2}$ defined by
\begin{equation}
\kappa _{\ell,j,m,n_1,n_2}=\sqrt{\mu +A_{\ell,j,m,n_1,n_2} \ }=\sqrt{\mu + \left(\ell+2j+2m+3n_1 + 3n_2+ d+3a_1+3a_2+\frac{11}{2}\right)^{2} \ }
\label{3cb-33}
\end{equation}
 The solution of the radial equation (\ref{3cb-32})  vanishing at $r=0$ and $r \to \infty$ are :

\begin{equation}
F_{k,\ell,j,m,n_1,n_2}(r)=r^{\kappa _{\ell,j,m,n_1,n_2}+\frac{1}{2}}\exp \left(-\frac{\omega r^{2}}{2}
\right)L_{k}^{(\kappa _{\ell,j,m,n_1,n_2})}(\omega r^{2}),\qquad k=0,1,2... \   \label{3cb-37}
\end{equation}
 in terms of the  generalized Laguerre polynomials and the associated eigenenergies are given by
\begin{equation}
E_{k,\ell,j,m,n_1,n_2}=2\omega (2k+\kappa _{\ell,j,m,n_1,n_2}+1),\qquad k=0,1,2....  \label{3cb-38}
\end{equation}

Collecting all pieces, we conclude  the physically acceptable solutions of 
the Schr\"odinger equation  (\ref{3cb-49}) to be given by  
\begin{eqnarray}
& & \Psi _{k,\ell,j,m,n_1,n_2}(r,\alpha,\theta, \beta, \varphi_1,\,\varphi_2 )
 = \ r^{\kappa _{\ell,j,m,n_1,n_2}-2} \ e^{-\frac{\omega r^2}{2}} \ \ L_{k}^{ (\kappa_{\ell,j,m,n_1,n_2)}}(\omega r^{2}) \nonumber\\
& &\times (\sin \alpha )^{2j+2m+3n_1+3n_2+d+3a_1+3a_2+\frac{7}{2}}C_{\ell}^{\left(2j+2m+3n_1+3n_2 +d+3a_1+3a_2 +\frac{11}{2} \right)}(\cos \alpha )  \nonumber \\
& &\times (\sin \theta )^{2m+3n_1+3n_2+3a_1+3a_2+3} (\epsilon_3 \cos \theta)^{d+\frac{1}{2}} P_{j}^{\left( 2m+3n_1+3n_2+3a_1+3a_2+4,d \right)
}(\cos 2 \theta )  \nonumber \\
 & & \times (\frac{1}{\sqrt{2}})  (\sin \beta )^{3n_1+3a_1+\frac{3}{2}} (\cos \beta)^{3n_2+3a_2+\frac{3}{2}} P_{m}^{\left( 3n_1+3a_1+\frac{3}{2}, 3n_2+3a_2+\frac{3}{2}  \right)}(\cos 2 \beta )  \nonumber \\
& & \times sgn(\sin 3\varphi_1)^{[(1-\epsilon)/2]}   \vert \sin 3\varphi_1 \vert^{a_1+\frac{1}{2}}C_{n_1}^{\left( a_1+\frac{1}{2}\right)}(\cos 3\varphi_1 ),\, \nonumber\\
 & &  \times sgn(\sin 3\varphi_2)^{[(1-\epsilon)/2]}   \vert \sin 3\varphi_2 \vert^{a_2+\frac{1}{2}}C_{n_2}^{\left( a_2+\frac{1}{2}\right)}(\cos 3\varphi_2 ),\,     \label{compb} 
\end{eqnarray}
 with $\epsilon=1 $ for bosons and  $\epsilon=-1 $ for fermions and 
\begin{eqnarray}
k &=&0,1,2,..., \ell=0,1,2,..., j=0,1,2,.., m=0,1,2,..., n_1=0,1,2,...,  n_2=0,1,2,...   \nonumber \\
%0 &\leq &\varphi \leq \frac{\pi }{3}\, \quad \frac{1-\epsilon_1}{2} \frac{\pi}{2}  \leq \theta \leq 
0 &  \leq &\varphi \leq \frac{\pi }{3}\ , 0  \leq \beta \leq \frac{\pi}{2}, 
 \frac{1-\epsilon_3}{2} \frac{\pi}{2}  \leq \theta \leq
\frac{3-\epsilon_3}{2} \frac{\pi }{2} , \epsilon_3=\pm 1, 
0  \leq \alpha \leq \pi \nonumber\\
& &  \quad a_1=\frac{1}{2}\sqrt{
1+2\lambda_1 } \ ,\quad   \quad a_2=\frac{1}{2}\sqrt{
1+2\lambda_2}\ ,  \quad d=\frac{1}{2} \sqrt{1 + \frac{2g}{3} \ }\nonumber    
\end{eqnarray}
 It has  to be noticed   that, for  Bose  statistics,  a $\delta$ pathology  occurs in (\ref{compb})  for 
 $a_j=1/2$ ($\lambda_j=0, j=1,2$) and $d=1/2$ ($g=0$)

The normalization constants $N_{k,\ell,j,m,n_1,n_2}$ can be  calculated from
\begin{eqnarray}
&& \int_{0}^{+\infty }r^{5}dr\int_{0}^{\pi }\sin^4 \alpha \  d\alpha 
\int_{(1-\epsilon_3) \pi/4}^{(3-\epsilon_3) \pi/4} \sin^3 \theta \  \ d\theta
\int_{0}^{\pi/2} \sin (2  \beta) \ d\beta \int_0^{\frac{\pi }{3}}d\varphi_1   \int_0^{\frac{\pi }{3}}d\varphi_2
\nonumber\\
 & &  \times 
\Psi _{k,\ell,j,m,n_1n_2}(r,\alpha,\theta ,\beta,\varphi_1,\varphi_2 )\Psi_{k^{\prime },\ell^{\prime },j^{\prime },m^{\prime },n_1^{\prime },n_2^{\prime }}
(r,\alpha,\theta ,\beta,\varphi_1,\varphi_2)  =\delta_{k,k^{\prime }}\delta _{\ell,\ell^{\prime }}\delta _{j,j^{\prime }}\delta _{m,m^{\prime }}  \delta _{n_1,n_1^{\prime }} \delta _{n_2,n_2^{\prime }} N_{k,\ell,j,m,n_1,n_2} \nonumber .
%\label{3ca.40.bis}
\end{eqnarray}%
As above, use is made  of the orthogonality properties of 
 Gegenbauer, Jacobi and Laguerre polynomials.

The eigenenergies are

\begin{equation}
E_{k,\ell,j,m,n_1,n_2}= 2 \omega \left(1 + 2 k + \sqrt{\mu + \left( \ell + 2 j + 2 m + 3 n_1 + 3 n_2 + d + 3 a_1 + 3 a_2 +
+ \frac{11}{2} \right)^2  } \right)
\end{equation}

Partial degeneracies are observed   i.e. all solutions for which $\ell + 2 j + 2 m + 3 n_1 + 3 n_2=N$ is satisfied, N, being fixed.

\section{Conclusions}

In this paper, we have proposed and exactly  solved few-body quantum problems of four, five and six
 particles in the $D=1$ dimensional space. The particles are confined in a mean field generated by harmonic 
 or attractive Coulomb-type forces. The interactions between particles are governed by two-body inverse square   Calogero potentials with both
translationally and  non-translationally invariant many-body forces,  for the three problems treated.
The difficulties encountered  concerning the explicit expression of the eigenfunctions, in problems concerned by  the pairwise Calogero interaction between each pair of particles, have been overcome  by restricting the number of two-body interactions. Our choice singularizes the interaction between  one cluster of three particles with one particle for the four-body problem. For the five-body problem , we singularize the interaction between two clusters of two and three particles respectively. Finally, for the six-body problem, we choose the interaction between two clusters
 each one formed by three particles.
 
 Appropriate coordinates transformations, lead to the solution of  the stationary Sch\"odinger equation,
  by providing the full regular eigensolutions with the eigenenergies spectrum for these few-body bounded systems.
 
  For the four-body problem we pay attention to the irregular solutions. 
The  coupling constants  domains for which the irregular solutions,   become square integrable are  determined. 
   Also for the four-body problem we have replaced the  harmonic confinement 
 by an attractive "Coulomb-type" interaction, and the solutions are given for the bound states only.
     This  study  can be extended straightforwardly
to the   five and six-body problems.

The present results suggest to extend the construction of exactly solvable models to larger systems of particles.

{\bf Acknowledgements } We thank  Dr. R.J. Lombard for fruitful discussions.
One of us (A.B.) is  very grateful to the Theory Group of the IPN Orsay for its kind hospitality
and to the university of Constantine {\bf 1} for financial support.

\end{document}